\documentclass[sigconf]{acmart}

\usepackage{times}%, mathptm}
\usepackage{moreverb}
\usepackage{graphicx}
\usepackage{mathrsfs}
\usepackage{bm}
\usepackage{url}
\usepackage{amsmath}
\usepackage{amssymb}

\usepackage{microtype}
\usepackage{graphicx}
\usepackage{subfigure}
\usepackage{balance}
\usepackage{booktabs} % for professional tables
\usepackage{amssymb}
\usepackage{amsthm}
\usepackage{array}
\usepackage{graphicx}
\usepackage{clrscode}
\usepackage{subfigure}
\usepackage{multirow}
\usepackage{multicol}
\usepackage{float}
\usepackage{color}
\usepackage{xcolor}
\usepackage{amsopn}
\usepackage{mathrsfs}
\usepackage{mathtools}
\usepackage{amsmath}
\usepackage{booktabs}
\usepackage{arydshln}
\usepackage{hyperref}
\usepackage{blkarray}
\usepackage{enumerate}
\usepackage{courier}
\usepackage{mathrsfs}
\usepackage{rotating}
\usepackage{bm}
\usepackage{wrapfig}
\usepackage{subfigure}
\usepackage{array}
\usepackage{ragged2e}
\usepackage{hyperref}
\usepackage{amsmath}
\usepackage{threeparttable}    
\usepackage[ruled,linesnumbered]{algorithm2e}

\SetKwComment{comment}{ $triangleright$ \ }{}
\theoremstyle{definition}

\theoremstyle{theorem}

\theoremstyle{proof}

\theoremstyle{remark}

\AtBeginDocument{%
  \providecommand\BibTeX{{%
    \normalfont B\kern-0.5em{\scshape i\kern-0.25em b}\kern-0.8em\TeX}}}

\copyrightyear{2022}
\acmYear{2022}
\setcopyright{acmcopyright}
\begin{document}
\newcommand{\xu}[1]{{\color{red} xu: #1}}

\title{Gumble Softmax For User Behavior Modeling}
\author{Weiqi Shao$^{1,}$, Xu Chen$^{1,*}$, Jiashu Zhao$^{3}$,Long Xia$^{2}$, Dawei Yin$^{2}$}\thanks{$*$ Corresponding author}
% \affiliation{$^1$Beijing Key Laboratory of Big Data Management and Analysis Methods} 
\affiliation{\{$^1$Gaoling School of Artificial Intelligence\} Renmin University of China, Beijing 100872, China} 
\affiliation{$^2$Baidu Inc} 
\affiliation{$^3$Department of Physics and Computer Science, Wilfrid Laurier University} 
\affiliation{shaoweiqi@ruc.edu.cn, successcx@gmail.com, long.phil.xia@gmail.com , yindawei@acm.org }

\begin{abstract}
Recently, 
sequential recommendation systems are important in solving the information overload in many online services.
Current methods in sequential recommendation focus on learning a fixed number of representations for each user at any time, with a single representation or multi-interest representations for the user.
However, when a user is exploring items on an e-commerce recommendation system, the number of this user's interests may change overtime (e.g. increase/reduce one interest), affected by the user's evolving self needs. 
Moreover, different users may have various number of interests. 
In this paper, we argue that it is meaningful to explore a personalized dynamic number of user interests, and learn a dynamic group of user interest representations accordingly.
We propose a Reinforced sequential model with dynamic number of interest representations for recommendation systems
(RDRSR).
Specifically, 
RDRSR is composed of a dynamic interest discriminator (DID) module and a dynamic interest allocator (DIA) module.
The DID module explores the number of a user's interests by learning the overall sequential characteristics with bi-directional self-attention and Gumble-Softmax.
The DIA module allocates the historical clicked items into a group of sub-sequences and constructs user's dynamic interest representations. 
We formalize the allocation problem in the form of Markov Decision Process(MDP), and sample an action from policy $\pi$ for each item to 
determine which sub-sequence it belongs to.
Additionally, experiments on the real-world datasets
demonstrates our model's effectiveness.
\end{abstract}
\maketitle

\begin{figure}[t]
\hypertarget{figure 1}{}
\centering
\setlength{\fboxrule}{0.pt}
\setlength{\fboxsep}{0.pt}
\fbox{
\includegraphics[width=0.90\linewidth]{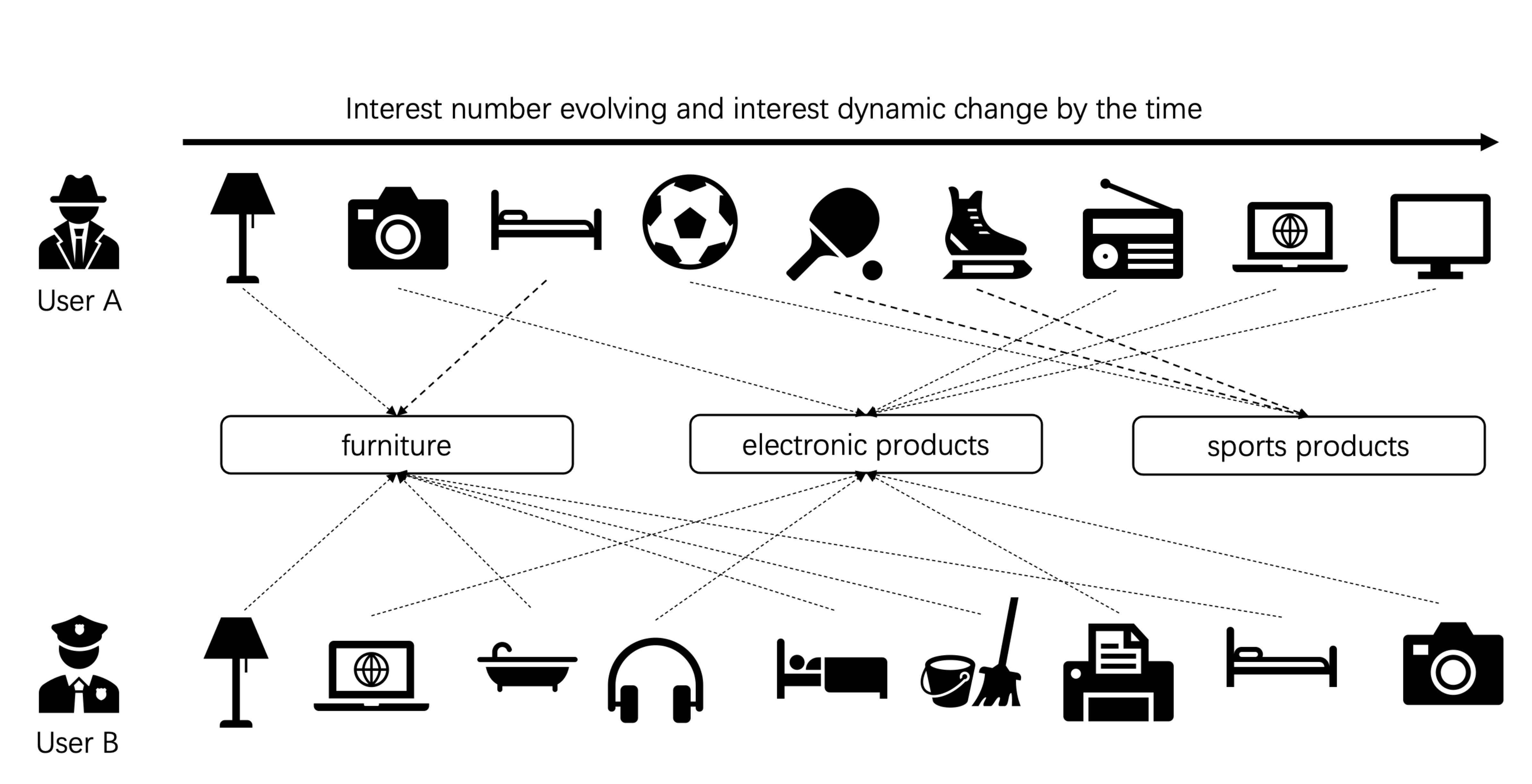}
}
\vspace*{-0.4cm}
\caption{
From the click sequences of user A
    and user B, there are multi conceptually distinct items 
    in a user’s click behaviors which indicates the change of dynamic number of user's interest and different interest number between users by the time.
}
\label{intro}
\vspace*{-0.4cm}
\end{figure}

\section{Introduction}
With the development of Internet technologies, recommender systems have been widely applied to many online services such as e-commerce, advertising, social media, and etc. %They serve as important information filtering technique to 
Recommender systems serve to alleviate the information overload problem and enhance user experiences. 
Traditional recommender systems mostly focus on promoting generalized user interests, such as collaborative filtering ~\cite{schafer2007collaborative, sarwar2001item}.
In recent years, more and more researchers study the sequential recommendation problem to capture the dynamic user behaviors, which assumes that a user's information need changes over the time \cite{rendle2010factorization}. 
%users’ intents are usually changing by the time. 
%It is vital to capture the dynamics of sequential user behaviors for making correct recommendations.

The existing sequential recommendation solutions represent a user as a fixed number of representations, including a single representation or multiple representations.
%include two types of approaches: a user is represente/  d as a single representation or multiple representations \cite{rendle2010factorizing,hidasi2015session,wang2015learning}. 
For the single representation recommendation, 
only one user embedding representation is generated for the next-item prediction.
Early solutions usually adapted the Markov Chain~\cite{rendle2010factorization} which assumes that the next-item prediction is closely related to the previous item~\cite{rendle2010factorizing}.
With the breakthrough of deep learning in many areas (e.g. computer vision and natural language processing)~\cite{zhang2019deep}, 
sequential neural networks such as recurrent neural network~\cite{hidasi2015session,li2017neural}and Transformer~\cite{vaswani2017attention}have been adopted to the sequential recommendation tasks. These sequential neural networks can characterize the sequential item interactions and learn informative representations for user behaviors~\cite{kang2018self}. Additional context information can also be considered to enhance the performance of neural sequential recommendation~\cite{zhang2019feature,huang2019taxonomy}.
%In the literature, different solutions~\cite{rendle2010factorizing}~\cite{hidasi2015session} ~\cite{wang2015learning}have been proposed for sequential recommendation systems,
%which including single-embedding method and multi-embedding method.
%In single-embedding methods, early solutions usually adapted the Markov which assumes that the next-item prediction is closely 
%related to the previous click item~\cite{rendle2010factorizing}.
%With the breakthrough of deep learning in computer vision and natural language processing~\cite{zhang2019deep}, 
%sequential neural networks such as recurrent neural network~\cite{hidasi2015session,li2017neural}and Transformer~\cite{vaswani2017attention}have been applied to recommendation tasks where these sequential networks can characterize the sequential item interactions and learn informative representations for user behaviors~\cite{kang2018self}.
%What's more, the context information has been also considered to enhance the performance of neural sequential recommendation like ~\cite{zhang2019feature,huang2019taxonomy}.
For multi-representation recommendation approaches, a user is assumed to have multiple interests and these interests jointly affect the user's next item selection. 
From the empirical analysis, a user usually interacts with several types of items that are conceptually different over time. For example, zhang ~\cite{zhang2019feature} identifies 
that the items in a user’s recent behaviors belong to different categories on Taobao dataset.
Various approaches have been adopted to model the multiple interests from the user's historical behaviors, including Capsule routing network~\cite{sabour2017dynamic} and multi-head self-attention~\cite{xiao2020deep}. The temporal information in the sequence can also be considered to enhance the recommendation performance~\cite{chen2021exploring}.
%Capsule routing network~\cite{sabour2017dynamic} has been used to extract the user's multi interest\cite{li2019multi,cen2020controllable} representation.
%Then ~\cite{xiao2020deep} explore user's multi representation with multi-head self-attentive, where the multi-head number as the multi-interest number through sum-pooling method.
%And other context information like time are considered in ~\cite{chen2021exploring}.
All the multi-interest modeling approach rely on a pre-given fixed number to generate the corresponding number of representations, which assumes that the numbers of interests for all users are the same and do not change over the time. 
%However, from the empirical analysis, a user 
%usually interacts with several 
%types of items that are conceptually different over time.
%Despite those single embedding or multi embedding methods have their own advantages and achieve significant performance improvements, all those methods generate an 
%fixed number user embedding representation for the next-item prediction.
%For example, zhang ~\cite{zhang2019feature} finds 
%that the items that belong to different categories 
%in a user’s recent behaviors on Taobao datase.

\begin{figure*}[h]
\hypertarget{figure 2}{}
    \centering
    \includegraphics[width=0.9\linewidth]{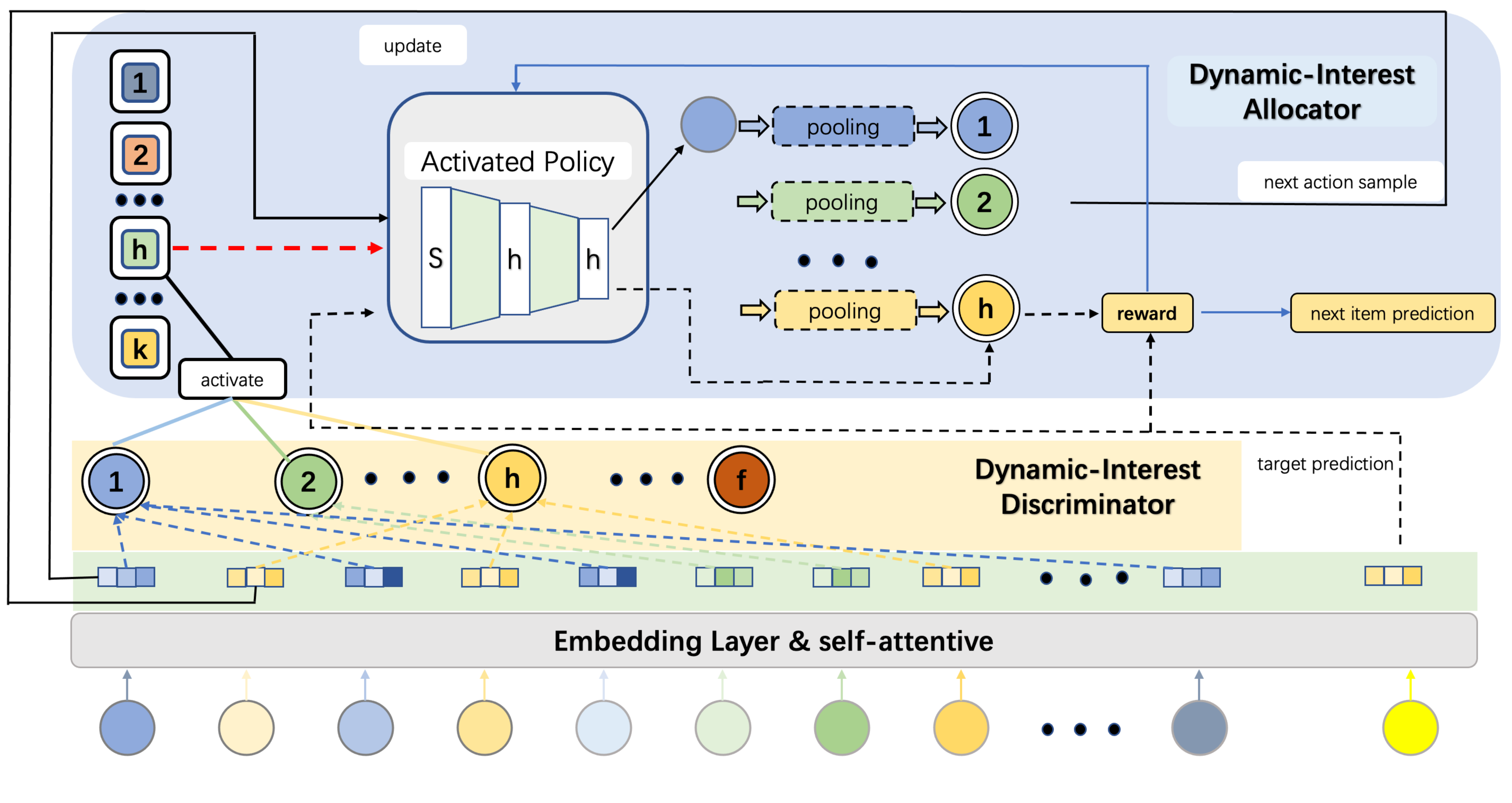}
    \caption{An overview of our model RDRSR. The input of our model is a user behavior sequence and those items are fed into the embedding layer and transformed into the item embeddings.
    Dynamic Interest Discriminator concentrates on exploring the dynamic interest number with a bi-directional architecture self-attention and Gumble-Softmax. Dynamic Interest Allocator activates the corresponding allocation policy
    according the learning interest number in DID.
    Then DIA allocates the click item into the sub-sequence with activeated policy and form different interests through average-pooling method.
    Last, we put the target item
    into the policy $\pi$ to decide which sub-sequence it
    belongs to and use the corresponding user interest representation
    to calculate the rewards between the interest and target item which will be used for prediction.}
    \label{fig:empunic1}
\end{figure*}

However, the fixed-number of interest assumption is not necessarily true in the real applications. For example, one user may have very broad interests, and another user have more focused intents. \hyperlink{figure 1}{Figure \ref{fig:empunic1}} shows two users each with a sequence of interacted (i.e. clicked) items, user A overall has three interests (furniture, electronic products, and sport products), while user B has two interests only ((furniture and electronic products). On the other hand, throughout the user behaviours over the time, a user may have more/less interests. In \hyperlink{figure 1}{Figure \ref{fig:empunic1}}, user A is only interested in furniture at the beginning, then A gradually start to show interest in electronic products and sport products. So we can see that user the number of A's interest changes from one 
to three. 
 Therefore, modeling a fixed number of interests can not fully simulate the real user intents. If a user has more interests than the given fixed number, then the user's intent can not be accurately represented. On the other hand, if a user has less interests than the given fixed number, then the user's intent will be represented with noise. Therefore, it is important to consider user's dynamic interest number in recommendation. In this paper, we propose a promising alternative method to learn a dynamic group of
embedding representations for a user’s behavior sequence, where each 
embedding representation encodes one aspect of the user’s intends.

%What's worse those recommendations generates a fixed number of representations for each user at any time neglect the user's dynamic interest number and different interest number between users.

Inspired by the above observations,
we introduce Learning Reinforced Dynamic Representations for Sequential Recommendation(RDRSR) to learn the a dynamic of group representations.
Specifically, we design Dynamic Interest Discriminator(DID) to detect the dynamic number of a user's interests
using self-attention~\cite{vaswani2017attention} and Gumble-Softmax~\cite{jang2016categorical}.
With self-attention, the items with high attention weights are 
clustered together to form different interests. And then with the informative item representations, Gumble-Softmax determines the interest number with the Gumble distribution as the noise to improve
the exploration of the user interest number.
Furthermore, 
%with the learning interest number and high cluster items representation, 
we design the Dynamic Interest Allocator(DIA)
to allocate the user's click sequence into a
dynamic group of interest sub-sequences, where
DIA formalizes the allocation process in the form of 
Markov Decision Process(MDP) and sample the action for 
each item to determine which sub-sequence it belongs to.
Here each sub-sequence forms a user's interest representation with average-pooling method.
As for the next-item prediction, we input the candidate item
into the policy $\pi$ to decide which sub-sequence it
belongs to and use the corresponding user interest representation
to calculate the compatibility between the sub-sequence and candidate item for prediction.

To summarize, the main contributions of this paper are:

\indent
$\bullet$ \
To the best of our knowledge, we are the first to consider a dynamic number of interest in
sequential recommendation.
The explore of user interest number 
improves the performance in the sequential recommendation.

\indent
$\bullet$ \
We propose the RDRSR model. The RDRSR model includes DID to learn the user's dynamic interest number over the time
and leverages the DIA to allocate the click into different sub-sequence to form
multi interests through average-pooling method for the next item prediction.

\indent
$\bullet$ \
We conducted experiments on several real datasets with several public benchmarks to verify the effectiveness of the model. We analyze the DID module and DIA module to valid the proposed RDRSR model through ablation study.

\section{Related work}
Before introducing the details of the proposed model, in this section, we introduce the related literature about recommendation systems, including general model, sequential model, multi-interest recommendation systems and attention mechanism we used in the paper.

\subsection{General recommendation}
The main methods in traditional recommendation system is extracting users’ general tastes from their historical behaviors to make recommendation. 
Typical methods include Collaborative Filtering~\cite{zhao2010user,sarwar2001item}, Matrix Factorization~\cite{koren2009matrix} and Factorization Machines.
Collaborative Filtering method is based on the similarity of users~\cite{zhao2010user} or the similarity of items~\cite{sarwar2001item} for recommendation. But it is a non-trivial work to quickly and accurately find the similar users or items.
Matrix Factorization(MF)~\cite{koren2009matrix} as one the most popular technique in recommendation system, map users and items into joint latent space and estimate user-item scores through the inner product between their embedding vectors. 
Factorization Machines(FM)~\cite{rendle2010factorization} methods consider all the variable interaction information which not only improve the recommendation results but also achieve good results even when the data is sparse.
With the success of deep learning in computer vision and natural language processing~\cite{zhang2019deep}, more and more efforts has been done to apply deep learning to the recommendation system~\cite{xu2018deep}.
He~\cite{he2017neural,he2016fast,he2017neural} makes a great success,
NCF~\cite{he2017neural}uses multi-layer perceptions to replace the inner product operation in MF for interaction estimation.
~\cite{he2016fast,he2017neural}use deep learning to obtain higher-oeder interactive expressions of interaction with a fast calculation trick.
These deep learning based methods achieve good performance.
Moreover, several attempts also tried to apply graph neural networks~\cite{fan2019graph,jin2020multi,tan2019deep}.

\subsection{Sequential recommendation}
In relevant literature,many sequential recommendation models
have been proposed to leverage user historical records in a sequential manner to capture the user’s preference for the next item.
By integrating the good performance of matrix factorization and the sequential pattern of Markov chains, factorized personalized Markov chains (FPMC)~\cite{rendle2010factorizing} embeds the sequential information between adjacent clicked items into the final prediction for recommendation, and later the hierarchical representation model (HRM)~\cite{wang2015learning} simultaneously consider the sequence behaviors and user preferences. Though they make progress in sequential recommendation, these methods only model the local sequential patterns between every two adjacent clicked item~\cite{yu2016dynamic}.
To model longer sequential behaviors, ~\cite{hidasi2015session} first adopted recurrent neural network to model the long sequence pattern for recommendation, RNN care too much about the sequence pattern which could be disturbed by the noise in the click sequence while neglect the user’s main intent, ~\cite{li2017neural,liu2018stamp} not only consider the sequence pattern in
the sequence and also explore the user’s main purpose through the
attention mechanism. Later, ~\cite{kang2018self} consider the importance of each item and other items in the click sequence achieve great progress in many real datasets~\cite{sun2019bert4rec} with unsupervised learning to learn the hidden relationships between items and make a difference.

\subsection{Multi-Interest recommendation systems}
The main difference between multi-interest recommendation and single embedding recommendation is that multi interest recommendation uses multi vectors to represent the user while only one vector in other methods.
The classic method~\cite{li2019multi,cen2020controllable} use a capsule routing based method to extract the user's multi interest. ~\cite{xiao2020deep}explore user's with multi-head self-attentive, where the multi-head number as the multi-interest number through sum-pooling method. 
~\cite{chen2021exploring}consider the time interval to extract the multi interest and ~\cite{tan2021sparse} infer a sparse set of concepts for each user from the large concept as its multi interest. 
Those methods have achieved good performance in recommendation, but non of them consider the different interest number between different users at different time and the dynamic user interest number over time.

\subsection{Attention}
The originality of attention mechanism is in computer vision~\cite{sun2003object,burt1988attention} to make the target object get more weight, but its great success in various fields in artificial intelligence comes only in recent years with the development of deep learning.
It first come to the center of the stage is in machine translation~\cite{bahdanau2014neural,vaswani2017attention} and is rather useful and efficient in real-world application tasks.
It is also been successfully applied in recommendation applications~\cite{xiao2017attentional}
which learns the importance of each feature interaction from data via a neural attention network. 
What's more, ~\cite{kang2018self,sun2019bert4rec} use the different relationships between items in the clicked sequence to capture both the long-term semantics and short-term semantics make a difference.

\section{Our model}
In this section, before going into the details of our proposed model.
We first describe the problem statement in our work. 
And then we will give an overview of the proposed
Learning Reinforced Dynamic Representations for Sequential Recommendation(RDRSR) framework(as shown
\hyperlink{figure 2}{Figure 2}),
which consists two main modules DID and DIA for dynamic interest number detector and 
user behavior allocation.

\subsection{Problem definition}
The key claim of sequential recommendation is that the current user preference should be related with the historical behaviors.
Formally, suppose we have a user set $\mathcal{U}$=$\{u_{1}, \!u_{2},...,\!u_{n}\}$, an item set $\mathcal{I}$=$\{i_{1}, \!i_{2},...,\!i_{m}\}$, 
and $n$ and $m$ are the numbers of users and items in the sequential recommendation task.
Unlike general recommendation, which only captures the correlation between a user and an item without considering the order of the click sequence.
We use $\mathcal{C}$=$\{x_{1}, \!x_{2},...,\!x_{t},\!{x_{t+1}}\}$ to denote a sequence of items in chronological order that a user has interacted, 
and the $x_i \in  \mathcal{I}$. 
The goal of sequential recommendation is to predict the next item $x_{|t+1|}$ depending on the precious click sequence $\{x_{1}, \!x_{2},...,\!x_{t}\}$.

\subsection{Embedding Layer}
 We create an item embedding matrix $\ E_{item} \in \mathcal{R^{m*d}}$ and an user embedding matrix $\ E_{user}\in \mathcal{R^{n*d}}$, where d is the latent dimension and $n$ and $m$ are the number of user and item.
 We retrieve the click item in the click sequence $\mathcal{C}$=$\{x_{1}, \!x_{2},...,\!x_{t}\}$ with a latent vector in the item embedding embedding and
 get the item sequence embedding $E_I$=$\{e_{1}, \!e_{2},...,\!e_{t}\}$ and the corresponding user embedding $e_{u}$, where t is the click sequence length and we process the datasets like~\cite{kang2018self}
 Furthermore, we incorporate a learnable position encoding matrix $P \in R^{txd}$ 
 to enhance the input representations. In this way, the input representations
 $E \in R^{t*d}$for the generator can be obtained by summing two embedding matrices:
 $E=E_I+P$.

\subsection{Dynamic Interest Discriminator}
As mentioned before, the user's
dynamic interest number is evolving and changing by the time,
a new click item would indicate user
get one more interest or
reduce a interest due to he may get what he want.
DID aims to find the user dynamic interest number
with the user's current click sequence.
First, we stack multiple bi-directional architecture self-attention~\cite{vaswani2017attention} block
based on the embedding layer.
With the bi-directional architecture self-attention block,
interest relevant items in the click sequence are clustering more close and 
get a more informative item representation.

\textbf{Self-Attention}
From the formula, the attention layer calculates a weighted sum of all values, where the weight between query and value, which could cluster those items belong to the same interest and effectively find the dynamic interest number.

{\setlength\abovedisplayskip{6pt}
\setlength\belowdisplayskip{6pt}
\begin{equation}
\begin{aligned}
Attention(Q, K, V) = softmax(\frac{QK^T}{\sqrt{d}})V
\end{aligned}\label{rand}
\end{equation}}

And the scale factor $\sqrt{d}$ is to avoid overly large values of the inner product when the dimension is very high.

We take E as input, convert it to three matrices through linear projections, and feed them into an attention layer:

{\setlength\abovedisplayskip{6pt}
\setlength\belowdisplayskip{6pt}
\begin{equation}
\begin{aligned}
\mathcal{S} = Attention(EW^Q, EW^K, EW^V)
\end{aligned}\label{rand}
\end{equation}}

where the projections matrices $W^Q$ $W^k$ $W^v \in \mathcal{R^{d*d}}$.
The projections make the model more flexible.

\textbf{Feed Forward Network}
Though the attention calculation is able to aggregate previous items’ embeddings with corresponding weights, 
it is still a linear model. In order to enforce the model with non-linearity and to get more high-order interaction information, we apply a two-layer feed-forward network to all $\mathcal{S}_i$.
\hypertarget{gongsi 3}{}
\begin{equation}
\begin{aligned}
F_{i} = FFN(\mathcal{S}_i) = ReLU(\mathcal{S}_iW^1+b^1)W^2+b^2
\end{aligned}\label{1}
\end{equation}
where $W^1$,$W^2$ are d x d matrices and $b^1$,$b^2$ are d-dimensional vectors.

In order to get the user's dynamic interest number,
we set an attention mechanism with the F and $e_u$ to get the
united user general purpose representation.
\begin{equation}
\begin{aligned}
a_{i} = Softmax((F_iW_{f1}+e_uW_u)W_{f2}+b)
\end{aligned}\label{1}
\end{equation}
\begin{equation}
\begin{aligned}
f = W_k(\sum_{i}^{t}a_{i}F_i)
\end{aligned}\label{1}
\end{equation}
where f is a k-dimension vector represent the 
probability of each possible interest number,
$W_k$ is a dxk matrix and k is the max dynamic interest number set in our model.

\textbf{Gumble Softmax Sampling}
We employ the Gumble Softmax~\cite{jang2016categorical} sampling method to produce the user dynamic interest number.
DID(Dynamic Interest Discriminator) draws z from a categorical distribution with class probabilities
$f=\{f_{1}, \!f_{2},...,\!f_{k}\}$. 
\hypertarget{gongsi 4}{}
\begin{equation}
\begin{aligned}
h = argmax_{i}[g_i + \log f_i]
\end{aligned}\label{1}
\end{equation}

where $h$ is current generated dynamic interest number and $\{g_{1}, \!g_{2},...,\!g_{k}\}$ are sample drawn from $Gumble$(0,1)
distributions. In practice, we sample the $Gumble$(0,1) distribution  using inverse transform sampling by drawing u from a uniform distribution.
What's more, those added new Gumble distribution as the noise changes the
probability distributions and give other original non-max alternative interest number
chance to be chosen, which
improve the exploration of the user interest number
and make our model more solid.
\hypertarget{gongsi 5}{}
\begin{equation}
\begin{aligned}
g = -\log(-\log(u))
\end{aligned}\label{1}
\end{equation}
where u is sampling from Uniform(0,1).

The argmax operation in \hyperlink{gongsi 4}{Eq. (6)} is non-differentiable, but we can resort to the Gumbel Softmax distribution,
which adopts softmax as a continuous relaxation to argmax in order to alleviate the non-differentiable problem used in \hyperlink{gongsi 11}{Eq. (11)(12)}. 
\hypertarget{gongsi 6}{}
\begin{equation}
\begin{aligned}
z_i = \frac{exp((\log f_i+g_i)/\mathcal{T})}{\sum_{j=1}^k exp((\log f_j+g_i)/\mathcal{T})} for\ i=1,2,\dots,k
\end{aligned}\label{1}
\end{equation}
where $\mathcal{T}$ is a temperature parameter to control the discreteness of the output vector $z$, which is set 10 in our model. 
Now, we get the probability $z$ for each dynamic interest number.
During the forward pass, we sample the dynamic interest number $h$ using \hyperlink{gongsi 4}{Eq. (6)} for the click item sequence. As for the backward pass, we are able to estimating the gradients of the discrete
samples by computing the gradients of the continuous softmax relaxation $z$ in \hyperlink{gongsi 6}{Eq. (8)}.

\subsection{Dynamic Interest Allocator}
After the user's dynamic interest number $h$ generated in \hyperlink{gongsi 4}{Eq. (6)}
and more informative representation $F$ in \hyperlink{gongsi 3}{Eq. (3)} of item vector found in DID,
DIA split the click sequence into different sub-sequences, where each 
sub-sequence represents a user's interest and
we use average-pooling method
to get the user's interest representation of those sub-sequence.
DIA formalizes the allocation click sequence problem in the form of Markov Decision Process(MDP), and sample action form policy $\pi$ for each item to determine which sub-sequence it belongs to.
Item representation $F$ with the bi-directional architecture self-attentive,
our policy $\pi$ can foresee future sequential information when making a decision, which could offer insightful clues to determine item-level relevance without direct supervision signals.
We consider an episodic RL approach to allocate the click sequence C${_i}$=$\{x_{1}, \!x_{2},...,\!x_{t}\}$ into h sub-sequence $S_{seq}$=$\{sub_{1}, \!sub_{2},...,\!sub_{h}\}$ and each sub-sequence represent a user interest represenation.

\textbf{Episode RL}
We see the allocation sequence split as an episode RL approach. 
At each time T, the process is in some state $s^T\in S$. According to the state $s^T$, the agent performs an action $a_i^T$ modeled by a policy $\pi(a_i^T|s^T)$. 
The action space is $a\in\{a_{1}, \!a_{2},...,\!a_{h}\}$,
where $a_i^T$ is that at time T, 
the $item^T$ is belongs to $sub-sequence_{a_i}$.
The following is the policy $\pi$.
\hypertarget{gongsi 9}{}
\begin{equation}
\begin{aligned}
\pi(a_i^T|s^T)= Softmax(ReLU(s{^T}W_{p1}+b_{p1}))W_{p2}+b_{p2}
\end{aligned}\label{1}
\end{equation}
where $\pi(a_i^T|s^T)$ 
is the discrete probability distribution that $item^T$ belongs to 
sub-sequence $a_i^T$ and
$W_{p1}$ is a d x $d_i$ matrix and $W_{p2}$ is a $d_i$ x h matrix.

\textbf{State Transition}
We give each sub-sequence an initial multi interest representation at time 0 $P^0$ = $\{p_{1}^0, \!p_{2}^0,...,\!p_{h}^0\}$ where each $p_{i}$ is a d-dimension vector and initialize with the corresponding user embedding $e_{u}$. At time T, we put the $s^{T}$ into the policy $\pi$ to get the action $a_i^T$(the sub-sequence $a_i$ $item^T$ belongs to).
We then use our well-designed pooling method to update the corresponding interest representation
embedding with the new added $item^T$, where the representation
of $item^T$ is $F_T$.
\begin{equation}
\begin{aligned}
p_i^{T+1}=average-Pooling(p_i^T,F_{T})
\end{aligned}\label{1}
\end{equation}
In reality, there are complex relationships between the user’s click sequence, like point level,union level with or without skip\cite{tang2018personalized}.
For accurately capturing those relationships,
we use a well-designed attention mechanism to define the state transition,
which explore the relationships between the new click item and the generated
sub-sequence with a weighted sum $p_i$, $F_{T+1}$  in \hyperlink{gongsi 3}{Eq. (3)}
and dynamic interest number distribution probability information $z$ in
\hyperlink{gongsi 6}{Eq. (8)}
through
a Neural Networks to get the $s_{T+1}$
\hypertarget{gongsi 11}{}
{\setlength\abovedisplayskip{6pt}
\setlength\belowdisplayskip{6pt}
\begin{equation}
\begin{aligned}
s_{T+1} = concat(\sum_{j=1}^{h}\alpha_{j}p_{j}^T,F_{T+1},z)W^0
\end{aligned}\label{rand}
\end{equation}}

{\setlength\abovedisplayskip{6pt}
\setlength\belowdisplayskip{6pt}
\begin{equation}
\begin{aligned}
\alpha_{j} = \frac{\exp((p_{j}^T\cdot F_{T+1}))}{\sum_{j=1}^{h} \exp((p_{j}^T\cdot F_{T+1}))}
\end{aligned}\label{rand}
\end{equation}}

where $W^0$ is a $2d x d$ matrix and $(\cdot)$ represent the inner product.

Even we use a hard allocate, but some information from other other sub-sequence is transitioning into the dynamic interest representation
when we define the state transition, which makes our model more solid.

\hypertarget{reward setting}{}
\textbf{Reward Setting}
After the allocation process, we get the multi interests representation at time t $P^t$ = $\{p_{1}^t, \!p_{2}^t,...,\!p_{h}^t\}$.
With the generated dynamic multi interests, 
here comes to the question that 
which interest representation is related to the target item.
To confirm the target interest,we use the target item $p_{target}^t$ to get the
current state $s^{t+1}$ through formulas $\hyperlink{gongsi 11}{Eq. (11)(12)}$ and put it
into the policy net $\pi(a_i^{t+1}|s^{t+1})$ set in \hyperlink{gongsi 9}{Eq. (9)} to sample action for getting the sub-sequence $p_{target}^t$ the target item belongs to.
Here we leverage a Sampled Softmax technique~\cite{covington2016deep,jean2014using} to calculate reward where the relationship $p_{target}^t$ with the target item and
other candidate item will be considered.
\hypertarget{gongsi 13}{}
\begin{equation}
\begin{aligned}
R_{c}=\frac{\exp((p_{target}^t\cdot e_{target}))}{\sum_{i=1}^{o}
\exp((p_{target}^t\cdot e_i))}
\end{aligned}\label{1}
\end{equation}
o is the sample item number in the dataset.
Through $R_c$ consider other items when calculate the reward,
it doesn't use other generated multi interests which means that
only when our target interest selection is correct, 
the reward is the optimal result.
In order to promote the policy $\pi$ to choose the right action, we employ a baseline in the reward function which use the average scores of the all generated
multi interests, defined as:
\begin{equation}
\begin{aligned}
R_{baseline}=\frac{\sum_{j}^{h}\frac{\exp((p_{j}^t\cdot e_{i}))}{\sum_{i=1}^{0} \exp((p_{j}^t\cdot e_i))}}{h}
\end{aligned}\label{1}
\end{equation}
with the baseline reward setting,
and the advantage of selected dynamic interest representation reward setting is as below:
\begin{equation}
\begin{aligned}
R_{advantage}=R_{c}-R_{baseline}
\end{aligned}\label{1}
\end{equation}
In order to enforce the learned dynamic multi interests representation orthogonally. 
Specific, we denote the $R_{orthogonal}$ as the mean of the absolute value
of the inner product between all different generated dynamic interest representations $p_i^t$ in $P^t$.
\begin{equation}
\begin{aligned}
R_{orthogonal}=-\frac{\sum_{i=1}^{h}\sum_{j=i+1}^{h}|p_i^t\cdot p_j^t|}{\frac{h*(h-1)}{2}}
\end{aligned}\label{1}
\end{equation}
where $|\cdot|$ represents the absolute value of inner product between $p_i^t$ and $p_j^t$ in $P^t$.
Combine the two reward above, 
the final reward function of our model is:
\hypertarget{gongsi 17}{}
\begin{equation}
\begin{aligned}
R_{s}=R_{advantage}+\lambda_o * R_{orthogonal}
\end{aligned}\label{1}
\end{equation}
where $\lambda_o$ is the trade-off parameter to balance the two rewards, which is set 0.001 in our experiments.

\subsection{Model optimization}
We treat the allocation task as a RL
problem and apply the classic policy gradient to learn the model
parameters. Specifically, the corresponding probability of generating $p_{target}^t$ sub-sequence is
$\mathcal{P}(sub)$ which is calculated as follows:
\begin{equation}
\begin{aligned}
\mathcal{P}(sub)=\prod_{T=1}^{t} \pi(a_i^T|s^T,\theta)*P(s^{T+1}|s^T,a_i^T,\theta)=\prod_{T=1}^{t} \pi(a_i^T|s^T,\theta)
\end{aligned}\label{1}
\end{equation}
The $p_{target}^t$ is then used for the dynamic interest selection for target item
$\pi(a_i^{t+1}|s^{t+1})$ in 
\hyperlink{reward setting}{Reward Setting}.
Thus, the probability of the generated sample action sequence is as followed:
\begin{equation}
\begin{aligned}
\mathcal{P}(s)=\mathcal{P}(sub)*\pi(a_i^{t+1}|s^{t+1})
\end{aligned}\label{1}
\end{equation}
Formally, the objective of the policy network is to maximize the expected reward
at the final prediction.
\begin{equation}
\begin{aligned}
\mathcal{J(\theta)}=E[R_i|\theta]=\sum_{s \in C} R_s * \mathcal{P}(s)
\end{aligned}\label{1}
\end{equation}
where $R_s$ is defined in \hyperlink{gongsi 17}
{Eq.(17)} and its gradient will be detached in the training process,
C is the all the generated action sequence of target sub-sequence and
$\theta$ is the parameters of the model including the parameters of DIA and DID.
The gradient of the objective function $\nabla_{\theta}\mathcal{J(\theta)}$
regard to the model parameters $\theta$ can derived as:
\begin{equation}
\begin{aligned}
\nabla_{\theta}\mathcal{J(\theta)}=\nabla_{\theta}\sum_{s \in C} R_s * \mathcal{P}(s)
\quad \quad \quad\quad\quad\quad\quad\quad
\\
=\sum_{s \in C}\nabla_{\theta}R_s * \mathcal{P}(s)
\quad \quad \quad\quad\quad\quad\quad\quad
\\
=\sum_{s \in C}\mathcal{P}(s)*R_s *\nabla_{\theta} \log(\mathcal{P}(s))
\quad \quad \quad
\\
=\sum_{s \in C} \sum_{T=1}^{t+1}\mathcal{P}(s)R_s *\nabla_{\theta} \log(\pi(a_i^T|s_T,\theta))
\\
=E_{s \in C}[\sum_{T=1}^{t+1}R_s *\nabla_{\theta} \log(\pi(a_i^T|s_T,\theta))]
\quad
\end{aligned}
\end{equation}
Therefore, the optimization of the policy network is calculate with a log trick as follow:
\begin{equation}
\begin{aligned}
\mathcal{L}_{RL}=-\log(\mathcal{P}(s))*R_s
\end{aligned}\label{1}
\end{equation}

Here we use the standard cross-entropy and a Sampled Softmax technique~\cite{covington2016deep,jean2014using} to calculate the classification loss:
\begin{equation}
\begin{aligned}
\mathcal{L}_{CE}=-\log \frac{\exp((p_{target}^t\cdot e_{target}))}{\sum_{i=1}^{o}
\exp((p_{target}^t\cdot e_i))}
\end{aligned}\label{1}
\end{equation}
o is the same as the negative sample number in \hyperlink{gongsi 13}{reward calculation}.
Finally, we jointly train the allocation task and 
classification task with a trade-off parameter $\beta$:
\begin{equation}
\begin{aligned}
\mathcal{L}=\mathcal{L}_{CE}+\beta * \mathcal{L}_{RL}
\end{aligned}\label{1}
\end{equation}
$\beta$ control the weight of the $\mathcal{L}_{RL}$ loss, which
is set 1 in our experiments.

\subsection{Prediction}
When we do the prediction, we first scan the user click session and 
select each action with the maximal probability at 
policy $\pi$ \hyperlink{gongsi 9}{Eq. (9)}
and corresponding state transition \hyperlink{gongsi 11}{Eq. (11)(12)}
, which can be written as follows:
\begin{equation}
\begin{aligned}
a_{max}^T = argmax_{a}{\pi(a_i^T|s^T)}
\end{aligned}\label{1}
\end{equation}
For each candidate item, we put it into policy net to get which sub-sequence it belongs to and get its corresponding reward.
We then rank all candidate items according to their rewards at 
\hyperlink{gongsi 13}{Eq. (13)} and return the top-$N$ rewards item as the final recommendations.

\section{Experiments}
In this section, we conduct experiments on sequential recommendation to evaluate the performance of our proposed method RDRSR on three benchmark
datasets 
compare with several state-of-the-art baselines. We first briefly introduce the datasets and the state-of-the-art
methods, then we conduct experimental analysis on the proposed model and the benchmark models.
Specifically, we try to answer the following questions:

\indent
$\bullet$ \
How effective is the proposed method compared to other state-of-the-art baselines?
$\mathbf{Q1}$

\indent
$\bullet$ \
What are the effects of the DIA(Dynamic Interest Allocator) and
DID(Dynamic Interest Discriminator)
modules through ablation studies? 
$\mathbf{Q2}$

\indent
$\bullet$ \
How sensitive are the hyper-parameter the max dynamic interest number $k$ in proposed model RDRSR?
$\mathbf{Q3}$

\subsection{Experimental Setup}
In this section, we introduce the details of the three experiment datasets, 
evaluation metrics, and comparing baselines in our experiments.

\textbf{Datasets}
We perform experiments on three publicly available dataset,
including MovieLens,
Lastfm
and Foursquare.
And the relative statistics information of the three datasets are shown in 
\hyperlink{table 1}{Table 1}.

\begin{table}[t]
\hyperlink{table 1}{}
\centering
\caption{\small{Statistics of the datasets.}}
\vspace{-0.3cm}
\small
\scalebox{.88}{
\begin{threeparttable} 
\begin{tabular}
{p{2.3cm}<{\centering}|
p{1.8cm}<{\centering}|
p{1.8cm}<{\centering}|
p{1.8cm}<{\centering}}
\toprule[1pt]
       \textbf{Dataset}     &\textbf{\# User}   &\textbf{\# Item}  &\textbf{\# Interaction} \\ \hline
{\textbf{MovieLens}}&944&1,683&100,000 \\
{\textbf{Foursquare}}&2,294&61,859&211,955 \\ 
{\textbf{Lastfm}}&1,860&2,824&583,933 \\ 

\bottomrule[1pt]

\end{tabular}
\end{threeparttable}  
}
\label{rec-dataset}
\vspace{-0.cm}
\end{table}

\begin{table*}[!t]
\hypertarget{table 2}{}
\caption{\small{Overall comparison between the baselines and our models. 
The best results are highlighted with bold fold. All the numbers in the table are percentage 
numbers with '\%' omitted.
}}
\center
\small
\renewcommand\arraystretch{1.05}
\vspace{-0.4cm}
\setlength{\tabcolsep}{5.1pt}
\begin{threeparttable}  
\scalebox{.87}{

\begin{tabular}
{p{0.8cm}<{\centering}p{0.8cm}<{\centering}
p{0.8cm}<{\centering}p{0.8cm}<{\centering}|
p{0.8cm}<{\centering}p{0.8cm}<{\centering}
p{0.8cm}<{\centering}p{0.8cm}<{\centering}|
p{0.8cm}<{\centering}p{0.8cm}<{\centering}
p{0.8cm}<{\centering}p{0.8cm}<{\centering}|
p{0.8cm}<{\centering}p{0.8cm}<{\centering}
p{0.8cm}<{\centering}p{0.8cm}<{\centering}
                      } \toprule[1pt]

\multicolumn{4}{c|}{} &
\multicolumn{4}{c|}{\textbf{Movielens}} &
\multicolumn{4}{c|}{\textbf{Foursquare}}&
\multicolumn{4}{c}{\textbf{Lastfm}}\\ 
 
\multicolumn{4}{c|}{} & 
\multicolumn{4}{c|}{Metric@10 \qquad Metric@50} &
\multicolumn{4}{c|}{Metric@10 \qquad Metric@50} &
\multicolumn{4}{c}{Metric@10 \qquad Metric@50} \\ \hline

\multicolumn{4}{c|}{\textbf{Single Embedding}} & 
\textbf{HR} & \textbf{NDCG} & \textbf{HR} & \textbf{NDCG} &
\textbf{HR} & \textbf{NDCG} & \textbf{HR} & \textbf{NDCG} &
\textbf{HR} & \textbf{NDCG} & \textbf{HR} & \textbf{NDCG} \\ \hline

\multicolumn{4}{c|}{\textbf{GRU4Rec}} & 
14.21 & 6.76 & 40.51 & 12.51 &
18.93 & 10.88 & 35.09 & 14.41 &
12.05 & 6.78 & 26.41 & 9.92 \\  

\multicolumn{4}{c|}{\textbf{STAMP}} & 
9.65 & 4.61 & 34.04 & 9.83 &
17.71 & 10.02 & 33.74 & 13.55 &
8.12 & 4.93 & 17.37 & 6.93 \\ 

\multicolumn{4}{c|}{\textbf{Caser}} & 
12.30 & 5.58 & 39.34 & 11.36 &
15.22 & 8.39 & 31.3 & 11.86 &
9.95 & 5.54 & 22.32 & 8.24 \\ 

\multicolumn{4}{c|}{\textbf{BERT4Rec}} & 
11.66 & 5.22 & 37.01 & 10.62 &
14.5 & 8.21 & 23.35 & 10.77 &
6.13 & 3.07 & 17.11 & 5.46 \\  \hline

\multicolumn{4}{c|}{\textbf{Multi Embedding}} & 
\textbf{HR} & \textbf{NDCG} & \textbf{HR} & \textbf{NDCG} &
\textbf{HR} & \textbf{NDCG} & \textbf{HR} & \textbf{NDCG} &
\textbf{HR} & \textbf{NDCG} & \textbf{HR} & \textbf{NDCG} \\ \hline

\multicolumn{4}{c|}{\textbf{MCPRN}} & 
14.42 & 6.51 & 42.74 & 12.64 &
19.15 & \textbf{10.91} & 37.49 & 14.94 &
12.05 & 6.33 & 27.43 & 9.65 \\ 

\multicolumn{4}{c|}{\textbf{RDRSR}} & 
\textbf{14.80} & \textbf{6.60} & \textbf{43.90} & \textbf{12.90} &
\textbf{20.40} & 10.50 & \textbf{41.90} & \textbf{15.30} &
\textbf{14.50} & \textbf{6.90} & \textbf{31.30} & \textbf{9.92} \\ 
\bottomrule[1pt]

\end{tabular}
}
        
\end{threeparttable}    
\label{tab:ab-result}   
\vspace{-0.3cm}
\end{table*}

$\mathbf{Ml-100k}$ \footnote{https://grouplens.org/datasets/movielens/100k/} is a dataset about 
user’s rating score for movies. In experiments, we follow ~\cite{he2016fast} to preprocess the dataset.

$\mathbf{Foursquare}$ \footnote{https://sites.google.com/site/yangdingqi/home/foursquare-dataset}
is a
location based social networks datasets which contains check-in, tip and 
tag data of restaurant venues in NYC collected from Foursquare from 24 October 2011 to 20 February 2012.

$\mathbf{Lastfm}$ \footnote{http://millionsongdataset.com/lastfm/}
records the music records of users from Last.fm. In experiments, we only use the click behaviors.

For Foursquare and Movielens datasets, we filter items and users interacted less ten times,
and five times in Lastfm datasets.
And all datasets are taken Leave-one-out method in~\cite{kang2018self}
to split the datasets into training, validation and testing sets. 
Specifically, we split the historical sequence for each user into three parts: (1) the most recent action for testing, (2) the second most recent action for validation, and (3) all remaining actions for training.
And if the click sequence length is less than t, we
repeatedly add a ‘padding’ item to the left until the length is
t.
Note that during testing, the input sequences contain training actions
and the validation actions.

\textbf{Baeslines}
We compare our proposed model RDRSR with the following state-of-the-art sequential recommendation baselines, including single representation methods and multi representation methods.

\textbf{Single representation models}
The most common sequential recommendation methods 
which generates a single embedding representation for the next-item prediction.

\indent
$\bullet$ \
\textbf{GRU4Rec}
~\cite{hidasi2015session} is a pioneering work which first leverages GRU to model user behavior sequences for prediction. 

\indent
$\bullet$ \
\textbf{Caser}~\cite{tang2018personalized}
is a recently proposed CNN-based method capturing sequential pattern by applying convolutional operations
on the embedding matrix for the most recent items,
and achieves state-of-the-art sequential recommendation
performance.

\indent
$\bullet$ \
\textbf{BERT4Rec}~\cite{sun2019bert4rec}
is a recently proposed BERT-based method which achieves state-of-the-art sequential recommendation
performance.

\indent
$\bullet$ \
\textbf{STAMP}~\cite{liu2018stamp} is a neural sequential model by incorporating user short-term memories and preferences.

\textbf{Multi representation model}
Sequential recommendation methods 
that generates multi representation to model 
user click behavior for the next-item prediction.

\indent
$\bullet$ \
\textbf{MCPRN}~\cite{wang2019modeling} is a recent representative work for extracting multiple interests which
designs a mixture-channel purpose
routing networks with a purpose routing network to detect the purposes of each item
and assign them into the corresponding channels to form multi presentations.

\textbf{Parameter Configuration.}
For a fair comparison, all baseline methods are implemented in Pytorch and optimized with Adam optimizer
with a mini-batch size of 2048. The learning rate is tuned in
the ranges of [0.01,0.001]. We tuned the parameters of comparing methods according to values suggested in original papers and set the embedding size $d$ as 64, and sequence length t=10. For our method, it has three crucial hyper-parameters: the trade-off parameter $\lambda_o$, $\lambda_x$ and the max dynamic interest number $k$.
We search $k$ from {3, 4, 5}, and we set $\lambda_o$, $\lambda_x$ 0.001 and 1.
In order to keep the policy consistent in Dynamic Interest Allocator,
we put the same user traing dataset in a batch to train the model.
The configuration of the other two parameters max dynamic interest number k and neg samples o for three 
datasets are reported in \hyperlink{table 3}{Table 3.}

\begin{table}[H]
\hyperlink{table 3}{}
\centering
\caption{\small{The optimal setting of our hyper-parameters for our model. Other parameters like dimension $d$ and learning rate $\gamma$ are set as 64 and 0.001, respectively.}}
\vspace{-0.3cm}
\small
\scalebox{.88}{
\begin{threeparttable} 
\begin{tabular}
{p{2.4cm}<{\centering}|
p{2.8cm}<{\centering}|
p{1.8cm}<{\centering}}
\toprule[1pt]

\multicolumn{1}{c|}{} &
\multicolumn{1}{c|}{\textbf{\ max dynamic interest number k}} &
\multicolumn{1}{c}{\textbf{\ neg samples o}}\\  \hline

{\textbf{MovieLens}}&4&99 \\
{\textbf{Foursquare}}&3&99 \\ 
{\textbf{Lastfm}}&3&199 \\ 

\bottomrule[1pt]

\end{tabular}
\end{threeparttable}  
}
\label{rec-dataset}
\end{table}

\textbf{Evaluation Metrics.}
For each user in the test set, we treat all the items that the user has not interacted with as negative items. We use two commonly used evaluation criteria ~\cite{he2017neural}: Hit Rate (HR) and Normalized Discounted Cumulative Gain (NDCG) to evaluate the performance of our model.

\subsection{Overall performance (Q1)}
\hyperlink{table 2}{Table 1}
 summarizes the performance of RDRSR and baselines including single-representation and multi-presenation methods on three benchmark datasets. Obviously, RDRSR achieves comparable performance to other the baselines on the evaluation metrics in general. 
 In the baselines of single representation methods, we find that
 GRU4Rec obtains good performance over other single-representation methods. 
 What's more, compare single representation methods with multi representation methods,
 it is obvious that recommendation with multiple presentations ( MCPRN, RDRSR) for a user click sequence perform generally better than those with single representation (Caser, GRU4Rec, BERT4Rec ...). Therefore, it is necessary to explore multiple representation to model user's diverse intents. 
 Moreover, we can observe that the improvement introduced by capturing user’s various intentions is more significant for Movielens and Lastfm datasets due to their density. 
 The users in denser datasets like Movielens and Lastfm tend to exhibit more diverse interests in online activity than rating datasets Movielens, which verifies the necessity of our motivation to model the dynamic interest number in the user behavior and the effectiveness of the DID module in exploring the user's dynamic interest number. 
 The improvement of RDRSR over the fixed interest number multi representation method(MCPRN) shows that dynamic interest exploration serves as a better multi-interest extractor than fixed multi interest.
 Considering the RDRSR and other baselines results, RDRSR consistently outperforms them on three datasets over all evaluation metrics. This can be attributed to two points: 1) The Dynamic Interest Discriminator explores user's dynamic interest number which takes the advantage of single representation methods when user's intent is one and multi representation methods when user's intents
 are more than one.
 2) RMRSR could correctly explore user's dynamic interest number and generates corresponding dynamic interest representation for next-item prediction
 while all other methods could be seen as fixed interest number method which are without enough flexibility.
 
 In our model, a major novelty is that we want to explore the user's dynamic interest number and form the corresponding dynamic interest representation. To obtain a better understanding why RDRSR performs better than other models, shown in 
 \hyperlink{figure 3}{Figure 3}, we further construct a case study on Movielens dataset. Specifically, we present a snapshot of the interaction sequence for a sampled user, which contains seven items,
 and top-one as the recommendatio result.
 Here we use different colors to represent the different dynamic interest sub-sequences, which is captured by the DID and DIA modules,
 and the total number of colors is equal to the user's dynamic interest number.
 The first five items are user's click behavior.
 In the first line at time t=6, the new movie doesn't not increase the user's dynamic interest number and the dynamic interest number is still 2.
 The second line at time t=6, the new movie with one more color yellow means that the user's dynamic interest number is increasing from
 2 to 3.
 Next, the user's new interest in sci-fi movie is main for the next-item prediction at time t=7.
 The result shows that our model can correctly explore the user's dynamic interest number and makes better recommendation.

% \begin{figure}
% \hypertarget{figure 3}{}
%     \centering
%     \includegraphics[width=\linewidth]{fig/paper4.jpg}
%     \caption{Case study.
%     The left before t=5 is the user behavior, at t=6 the user clicks
%     two different movies and get different recommendation results. Here we use colors to
%     represent dynamic interest sub-sequences.
%     The picture of each movie is downloaded from https://movie.douban.com.
    
%     }
%     \label{fig:empunic1}
% \end{figure}

\begin{figure}[t]
\hypertarget{figure 3}{}
\centering
\setlength{\fboxrule}{0.pt}
\setlength{\fboxsep}{0.pt}
\fbox{
\includegraphics[width=\linewidth]{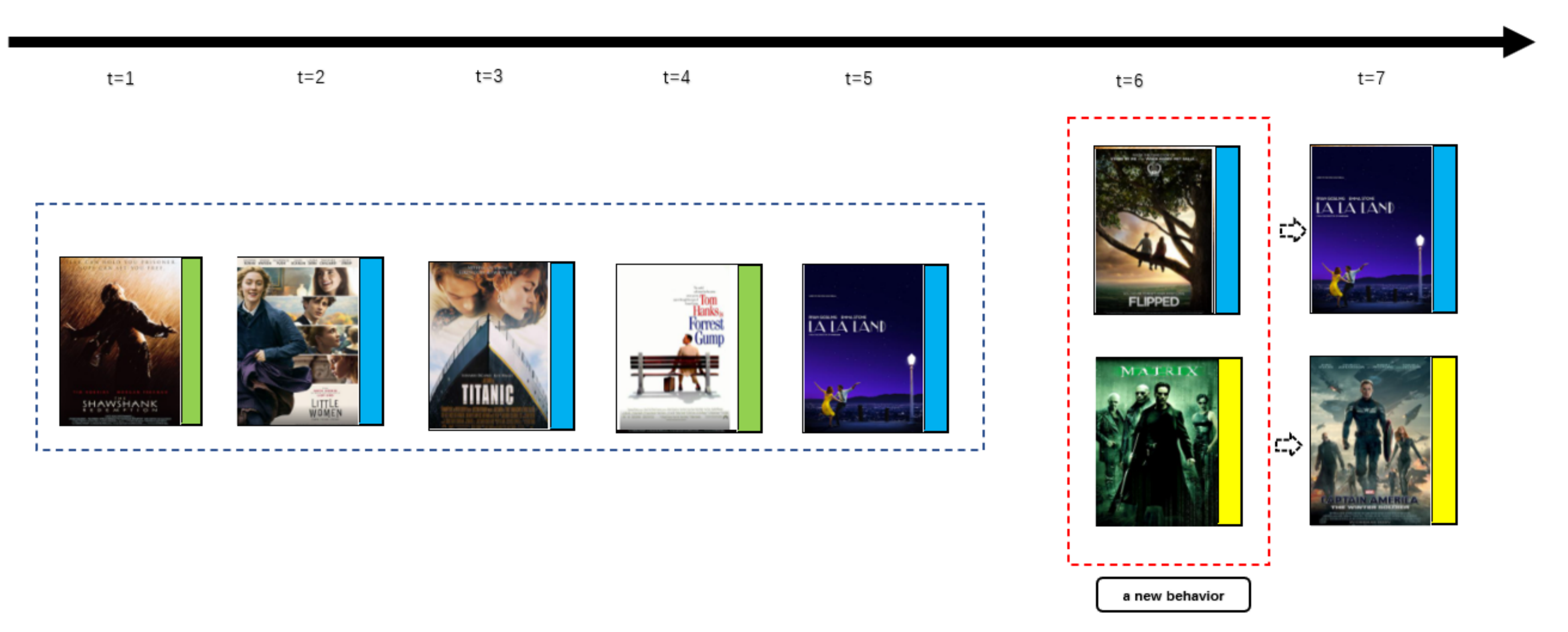}
}
\vspace*{-0.4cm}
\caption{
Case study.
    The left before t=5 is the user behavior, at t=6 the user clicks
    two different movies and get different recommendation results. Here we use colors to
    represent dynamic interest sub-sequences.
    The picture of each movie is downloaded from https://movie.douban.com.
}
\label{intro}
\vspace*{-0.4cm}
\end{figure}

\subsection{Ablation study (Q2)}
We introduce one variant (RDRSR-F) to validate
the effectiveness of the proposed model. 
Specifically, RDRSR-F shuts down the module Dynamic Interest Discriminator, 
and the the module Dynamic Interest Allocator set a fixed dynamic interest number.
We conduct
experiments on all three datasets.
\hyperlink{table 4}{Table 4} reports the results in
terms of NDCG@10. 
RDRSR-F3 and MCPRN-3 means that the fixed interest number is 3 and the max dynamic interest number
of RDRSR-3 is 3. 
Obviously, RDRSR-3 significantly outperforms the
variant RDRSR-F3 on all datasets. The substantial difference between
RDRSR-F3 and RDRSR-3  shows that the learning dynamic user dynamic interest number in DID module is better
than those fixed interest number in RDRSR-F3. And it verifies our
motivation to explore the user dynamic interest number in sequential recommendation and the effectiveness of the proposed module DID.
What's more, the improvement of RDRSR-F3 over MCPRN-3 validates that our DIA module is useful to model user's dynamic interest representations for next-item recommendation.
\begin{table}[H]
\hypertarget{table 4}{}
\centering
\caption{\small{Ablation study. 
Performance comparison of RDRSR-3 (max dynamic interest number 3), its variant RDRSR-F3 and MCPRN-3 (fixed interest number 3) over three datasets.
And all the numbers in the table are percentage 
numbers with '\%' omitted.}}
\vspace{-0.3cm}
\small
\scalebox{.88}{
\begin{threeparttable} 
\begin{tabular}
{p{1.6cm}<{\centering}|
p{1.6cm}<{\centering}|
p{1.6cm}<{\centering}|
p{1.6cm}<{\centering}|
p{1.6cm}<{\centering}}
\toprule[1pt]

\multicolumn{1}{c|}{\textbf{Datasets}} &
\multicolumn{1}{c|}{\textbf{\ Metric}} &
\multicolumn{1}{c|}{\textbf{\ MCPRN-3}} &
\multicolumn{1}{c|}{\textbf{\ RDRSR-F3}} &
\multicolumn{1}{c}{\textbf{\ RDRSR-3}}\\  \hline

{\textbf{MovieLens}}&{\textbf{NDCG@10}}&6.32&6.30&6.60 \\
{\textbf{Foursquare}}&{\textbf{NDCG@10}}&9.73&10.10&10.50 \\ 
{\textbf{Lastfm}}&{\textbf{NDCG@10}}&6.32&6.40&6.90 \\ 

\bottomrule[1pt]

\end{tabular}
\end{threeparttable}  
}
\label{rec-dataset}
\end{table}

\subsection{Hyperparameter study (Q3)}
We also investigate the sensitivity of
the max dynamic interest number $k$ to RDRSR in all three datasets. 
\hyperlink{figure 4}{Figure 4}
reports the performance of our model in the metrics of HR and NDCG. 
In particular,
We keep the other parameters in the model consistent with the Q1 settings. 
From the figure, we
can observe that RDRSR obtains the best performance of HR and NDCG when $k$ equals 3 or 4. 
With the fixed sequence length t the result 
increases with the increase of
the max dynamic interest number $k$, 
which indicates that the user's dynamic interest number is multi and bigger max dynamic interest number set in the model may meet the requirements better.
RDRSR becomes a single representation method(RDRSR-1) when the max dynamic interest number is 1.
The sub-optimal results achieved by RDRSR-1 gives evidences that 
single representation is not the best solution for sequential recommendation and the necessity of dynamic interest representations
methods.
The recommendation performance increases at the beginning, 
but decreases after reaching a peak due to the complex model structures with bigger max dynamic interest number k, which brings more noise
and makes sub-optimal recommendation.

\begin{figure}[]
\hypertarget{figure 4}{}
\centering
\setlength{\fboxrule}{0.pt}
\setlength{\fboxsep}{0.pt}
\fbox{
\includegraphics[width=0.90\linewidth]{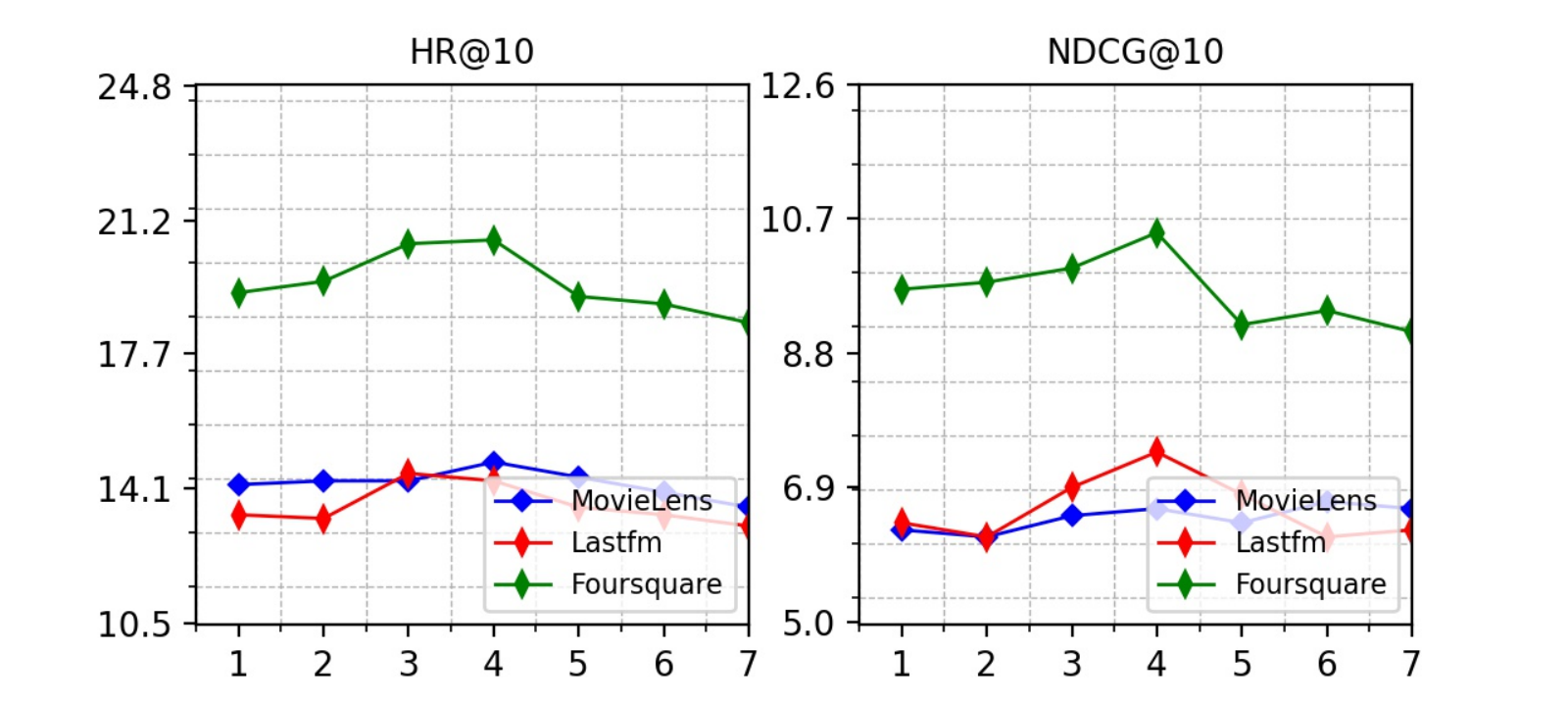}
}
\vspace*{-0.4cm}
\caption{
Hyperparameter study, where the 
    horizontal coordinates is the max dynamic interest number from 1 to 7,
    and the the vertical coordinates are the metric HR@10 and NDCG@10.
}
\label{intro}
\vspace*{-0.4cm}
\end{figure}

\section{Conclusion}
In this article,
we learning a dynamic group of representations for user to improve the performance of the sequential recommender system.
In order to achieve this goal, we design DID and DIA to capture
the dynamic interest number and form the corresponding dynamic interest representations. 
What's more, 
we conducted a ablation study to explore the effectiveness of DID and DIA modules and verified the effectiveness of RDRSR on several real datasets with SOTA methods. 
To the best of our knowledge, we are the first to consider the personalized dynamic interest number in sequential recommendation.
However,
the proposed model also exists shortcomings in computing speed,
where we formulate the allocation task in DIA module as a MDP problem which is computing cost and unstable in training.
In the future we will consider how to allocate the click sequence in a more effective way.

\bibliographystyle{ACM-Reference-Format}
\balance
\bibliography{main}

\end{document}